# A Conversation with Shoutir Kishore Chatterjee

**Tathagata Banerjee and Rahul Mukerjee**

*Abstract.* Shoutir Kishore Chatterjee was born in Ranchi, a small hill station in India, on November 6, 1934. He received his B.Sc. in statistics from the Presidency College, Calcutta, in 1954, and M.Sc. and Ph.D. degrees in statistics from the University of Calcutta in 1956 and 1962, respectively. He was appointed a lecturer in the Department of Statistics, University of Calcutta, in 1960 and was a member of its faculty until his retirement as a professor in 1997. Indeed, from the 1970s he steered the teaching and research activities of the department for the next three decades. Professor Chatterjee was the National Lecturer in Statistics (1985–1986) of the University Grants Commission, India, the President of the Section of Statistics of the Indian Science Congress (1989) and an Emeritus Scientist (1997–2000) of the Council of Scientific and Industrial Research, India.

Professor Chatterjee, affectionately known as SKC to his students and admirers, is a truly exceptional person who embodies the spirit of eternal India. He firmly believes that "fulfillment in man's life does not come from amassing a lot of money, after the threshold of what is required for achieving a decent living is crossed. It does not come even from peer recognition for intellectual achievements. Of course, one has to work and toil a lot before one realizes these facts."

SKC is a scholar and researcher of the highest order of eminence. His research and other contributions exhibit an amazing depth and cover such diverse areas as sequential analysis, nonparametric methods, design of experiments and foundational issues. Much in contrast with what seems to be the current practice, he has been extremely parsimonious in publishing his work. Only his very best findings, all of which require a lot of training and background to be fully appreciated, are available in print.

The department with which SKC was associated for more than three decades happens to be the oldest full-fledged postgraduate department of statistics in Asia. It was founded in July, 1941, with Professor P. C. Mahalanobis as the honorary head and R. C. Bose and S. N. Roy as full-time lecturers. Opening up this department is considered a momentous event in the development of statistics in India. Conversation with SKC also brings out the history of this department and its pioneering role in the advancement of statistics teaching and research in this part of the world.

The following conversation took place at the home of Professor Chatterjee in Calcutta on July 2, 2006.

*Tathagata Banerjee is Associate Professor, Indian Institute of Management Ahmedabad, Ahmedabad 380 015, India e-mail: tathagata@iimahd.ernet.in and is on leave from University of Calcutta. Rahul Mukerjee is Professor, Indian Institute of Management Calcutta, Calcutta 700 104, India e-mail: rmuk@iimcal.ac.in.*

## EARLY YEARS: LANDING IN STATISTICS

**Banerjee:** Please tell us about your family background and early life.

**Chatterjee:** I was brought up in Burdwan, at that time a small town, situated about sixty miles away from Calcutta, in a middle-class family which was not very well-to-do but had deep cultural moorings. My parents had literary interests as well as an appreciation for mathematics which, in the case of my mother, can be traced back to my maternal grandfather, who was a brilliant scholar both in mathematics and Sanskrit. Because of my family influence, I developed some interest in literature early in my life.

**Mukerjee:** What about mathematics? Which branches of mathematics interested you the most at high school?

**Chatterjee:** I had interest in geometry and algebra. Perhaps I was not strong enough in mathematical manipulations, but the reasoning part of the subject used to appeal to me.

**Mukerjee:** We know that the idealistic ambience of pre- and immediate post-independence India inculcated in you, like many of your generation, a sense of integrity and sincerity. Was there also any focus on ingenuity at the high school level?

**Chatterjee:** My parents and some of the teachers in my school used to put a lot of emphasis on originality of thought, both in literary composition and mathematical derivation.

**Banerjee:** How did you get interested in the subject of statistics?

**Chatterjee:** I used to find interest in most subjects, both literary and scientific, that I learned in school and during the first two undergraduate college years. Perhaps it would be correct to say that subjects which are conceptual and give free play to one's imagination held more attraction for me. Although mathematical reasoning attracted me, I cannot say that I had any special leaning toward mathematics. But I was rather clumsy in laboratory work, particularly in chemistry. In those days, the subject of statistics was little known outside a small knowledgeable circle. However, it was the only subject in the Bachelor of Science course which one could study at the honors (major) level without being forced to choose chemistry as one of the subsidiary subjects. This led me to get interested in statistics in the first place. (Interestingly, I had illustrious predecessors among statisticians with respect to lack of proficiency in chemistry laboratory work. Professor J. Roy once told me that he had come to statistics for precisely the same reason; also I have heard that Professor S. N. Roy had fared badly in the chemistry practical and suffered on that account.) Besides, whatever additional information about the subject of statistics I could gather pointed to its being an emerging discipline and holding out good prospects for the student.

**Mukerjee:** So you enrolled yourself at the Presidency College, Calcutta, for a bachelor's degree, with honors in statistics, and mathematics and physics as subsidiary subjects, and after getting your degree from there, joined the University of Calcutta for your master's. We know Presidency College, Calcutta (originally called Hindu College), founded in 1816, is among the premier institutions for western education in India. The University of Calcutta, founded in 1857, is again one of the oldest universities in India imparting western education. We also know that several internationally famous personalities both in the humanities and the sciences were associated with these venerable institutions as student, researcher or faculty. Tell us about the atmosphere there at that time.

**Chatterjee:** Presidency College (which is an affiliated college under the University of Calcutta) at that time attracted the best students from this part of the country and it had also one of the best faculties in the country. The general tendency among the better students to opt for engineering and medical courses, which set in a few years later, was still not there. Also, statistics was a very inviting subject at that time; so there was great competition among the mathematically minded students to get enrolled for the honors course in statistics. At the University of Calcutta too, bright students from different parts of the country flocked to enroll themselves for the master's degree in statistics. The atmosphere was very competitive, though friendly, and the intellectual standard was quite high.

**Banerjee:** How was the method of instruction at these institutions?

**Chatterjee:** The subject of statistics was then in a formative stage. There were very few textbooks that could be followed as course material. We had to depend on the lectures of our teachers, sometimes supplemented by the study of original research articles. Since the subject of statistics itself had not assumed a rigid contour in those days, there was a lot of freedom. It was all very exciting and the joy of learning was abounding.

**Banerjee:** Who were the most influential teachers?

**Chatterjee:** I must name Professor B. N. Ghosh at the Presidency College. What attracted me most was that he always approached a problem from the conceptual, rather than the formal mathematical, point of view. Professors M. N. Ghosh and H. K. Nandi were the most influential teachers at the master's level. Nandi used to take the major burden of the teaching load and had to teach a wide variety of subjects, but still his insightful and incisive remarks were always illuminating. M. N. Ghosh used to teach mathematical subjects. He never came prepared specifically for the class, but treated the topic to be taught as a sort of research problem and took the students along with him in reaching a solution.

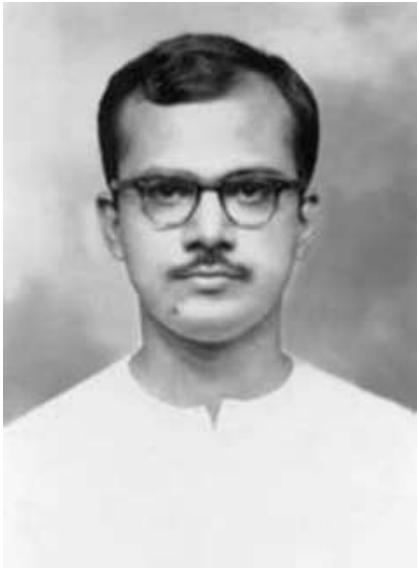

FIG. 1. *SKC as a doctoral student, 1958.*

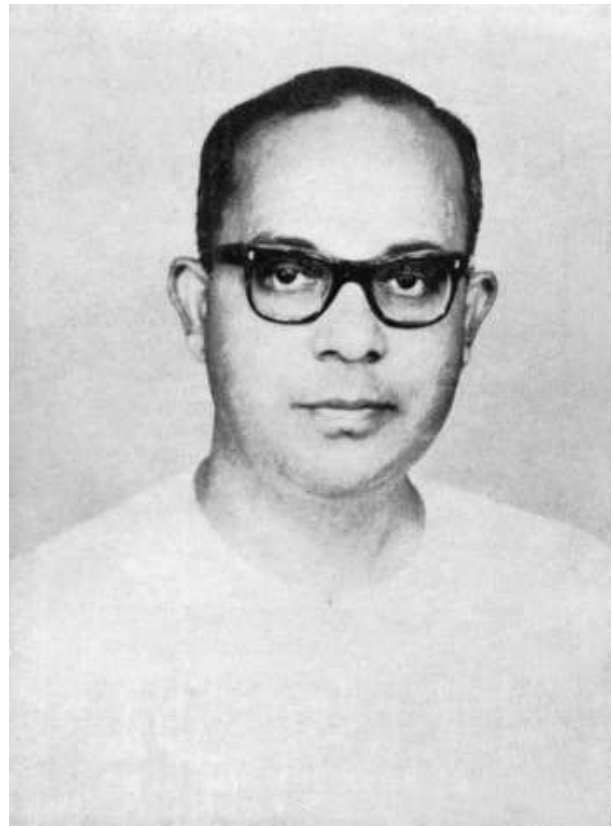

FIG. 2. *H. K. Nandi, the spirit behind the Calcutta school.*

**Banerjee:** We know that you had some interest in Indian classical music from your early college days. How did it develop?

**Chatterjee:** Actually, some of my classmates were remarkable exponents of such music. Whatever little interest in Indian classical music I developed (but it was never much) was under their influence. Incidentally, my teachers H. K. Nandi and M. N. Ghosh were both connoisseurs of classical music.

## AS A DOCTORAL STUDENT

**Mukerjee:** Why did you join a doctoral program at the University of Calcutta upon the completion of your master's degree?

**Chatterjee:** My well-wishers and I had by that time realized that any nonacademic profession would not suit my mental makeup. When I asked my father about joining a research career, he whole-heartedly supported me, even though it meant that with the small stipend that I would get, I would be able to contribute very little to the family coffers for the next three years or so. At that time my two brothers and my sister were already in for higher education and my father's resources were under some strain.

**Mukerjee:** In those days, many of the bright students opting for a research career initially joined the Indian Statistical Institute (ISI) and then used it as a springboard to land in universities abroad. Did you ever think of doing so?

**Chatterjee:** I straightaway joined the Department of Statistics, University of Calcutta, and never thought of joining ISI and going abroad later on.

**Banerjee:** How did you choose the topic of your doctoral research?

**Chatterjee:** I was interested in the area of statistical inference. My Ph.D. thesis concerned the development of sequential procedures of Stein's type with nuisance-parameter-free performance in the multivariate setup. Professor H. K. Nandi suggested the topic.

**Banerjee:** Please tell us about your interaction with Professor Nandi as your research adviser.

**Chatterjee:** Professor Nandi was a unique personality who allowed the student to take up a problem and then left him to fend for himself, typically with the barest suggestion or hint. But once the student picked up an original idea, Nandi used to act as a facilitator by offering help with incisive comments and suggestions on further work. Personally, I groped for about two years before making any real headway. After a good deal of probing of the approaches to multivariate extension, it struck me that instead of constructing a linear function of the observations on each variable separately, one should try to construct a set of linear functions of the observations on all the variables simultaneously. The rest of the work was completed in about one year [1, 2, 3].

**Mukerjee:** At the doctoral level, who were your contemporaries in the department?

**Chatterjee:** I joined as a research scholar in early 1957. P. K. Bhattacharya, who later on was at the University of California, Davis, was then nearing the completion of his thesis under the supervision of Nandi. In the next two years, successively Pranab (P. K. Sen, now at the University of North Carolina at Chapel Hill) and Jayanta (J. K. Ghosh, now at Purdue University) joined the department as research scholars, both under the supervision of Nandi. Later we three started teaching in the department in successive years.

## ABOUT THE DEPARTMENT

**Banerjee:** By this time, you must have had a thorough idea about the Department of Statistics, University of Calcutta. We know that this is the oldest full-fledged statistics department at the postgraduate level in Asia. Please recount in some detail the history of this department and its impact on the development of our subject in India.

**Chatterjee:** The department came into existence in the year 1941. At that time there were very few universities in the world with full-fledged departments devoted to teaching and research in statistics. In fact, the subject itself had a sort of amorphous identity. Certain parts of it were studied as probability theory and its applications, certain others in the context of agricultural or social sciences, and still others by actuaries in their own ways. It was to the credit of Professor P. C. Mahalanobis that, almost single-handedly, he was able to give the subject a footing in the University of Calcutta at that early stage.

**Banerjee:** How did the subject develop in India prior to the founding of the department?

**Chatterjee:** Mahalanobis in his youth had gone to study physics in England and, under the influence of the biometric school of Karl Pearson, became convinced of the potentialities of statistics. Although on his return to India he had joined the Presidency College as a professor of physics, he had taken the application and propagation of statistics in India as his life's mission. Around 1930, he had established in the premises of Presidency College a small unit for conducting theoretical and applied research in statistics; this later developed into the ISI. Toward the end of the 1930s, with the active involvement of scholars like R. C. Bose, S. N. Roy, A. Bhattacharya, B. N. Ghosh, P. K. Bose and others, who had been drawn to statistics from other disciplines such as mathematics and physics, the work of research and training in this unit had gathered considerable momentum.

**Mukerjee:** Why and how did Mahalanobis persuade the university authorities to establish a separate department of statistics?

**Chatterjee:** It was felt that without the creation of such a department, the pace of development of the subject in India would remain tardy. Mahalanobis with his persuasive skill was able to convince the authorities of the University of Calcutta about the need of such a department. I have heard that in one meeting of the university senate where the proposal was mooted, many members expressed their doubts about its viability, as the subject did not even have enough books to base a course upon. In the next meeting, Mahalanobis carried on the heads of porters several basketloads of books and journals and forced a favorable decision. Thus in 1941, a new master's level course in statistics got started in the university, naturally with students who at the bachelor's level had honors in other disciplines and therefore practically no previous exposure to the subject. This shortcoming was remedied three years later when a bachelor's level honors course in statistics was started in Presidency College.

**Mukerjee:** What was the impact elsewhere in India of the founding of the department in the University of Calcutta?

**Chatterjee:** Within a few years, it inspired other universities of the subcontinent such as, to name the earlier ones, those at Trivandrum, Patna, Gauhati, Dacca, Bombay, Lucknow, Pune and Banaras, to start similar departments. They generally adopted variants of the statistics curriculum of our department according to their situations. Besides, in the 1940s a strong center developed in the Indian Agricultural Statistics Research Institute in New Delhi under the leadership of P. V. Sukhatme. I guess, at present statistics is taught at the master's level in about sixty centers in India.

**Mukerjee:** What was the structure of the department in those early days?

**Chatterjee:** Mahalanobis was the honorary Head of the new department. The faculty included R. C. Bose, S. N. Roy, A. Bhattacharya and others. A little later, B. N. Ghosh and P. K. Bose also joined the faculty as part-timers. The first batch of students enrolled included C. R. Rao and H. K. Nandi. Both of them joined the department as faculty immediately after obtaining their master's degrees in statistics two years later and so did M. N. Ghosh, who had his degree in pure mathematics. But it should be mentioned that all these people worked concurrently in ISI. A few, like R. C. Bose and S. N. Roy, had substantive appointment in the university, but most of them were whole-timers of ISI and taught in the university as guest teachers. In fact, in the initial years ISI and the university department were so organically associated with each other that it would have been difficult to draw a line between the activities of the two.

**Banerjee:** How did the department start assuming an identity separate from ISI?

**Chatterjee:** The department, as part of a heritage university, had the responsibility of consolidating teaching and academic research in statistics. On the other hand, Mahalanobis wanted ISI to promote the cause of the subject by exploring all possible channels and, in particular, by conducting large-scale sample surveys on behalf of the government. Inevitably, there was a parting of ways.

**Banerjee:** Was there any significant intellectual cooperation between the department and ISI at this stage?

**Chatterjee:** It continued, albeit at a low ebb. At that time, ISI did not have the authority to confer degrees. Therefore, for some years, researchers there had to get themselves enrolled in the University of Calcutta for their Ph.D. degrees.

**Banerjee:** What were the initial challenges after the department decided to chart out an independent course for itself?

**Chatterjee:** A severe jolt came around 1950, when first R. C. Bose and thereafter S. N. Roy, who had successively headed the department after

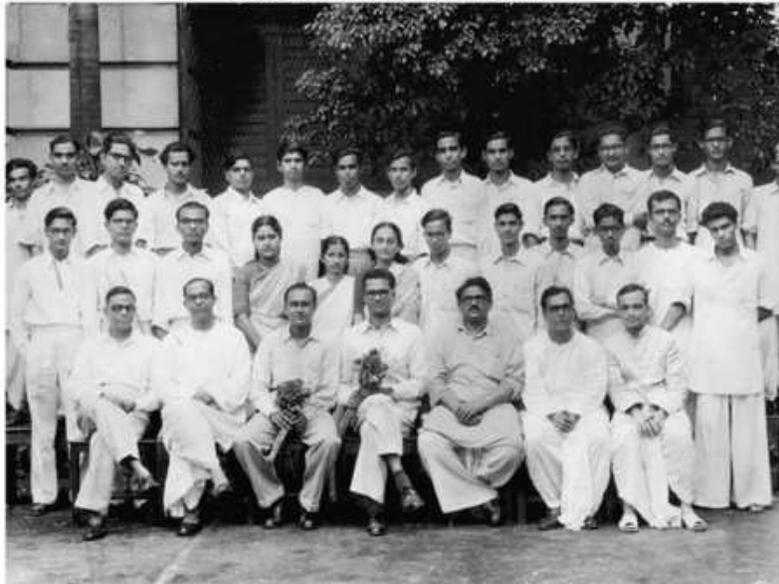

Fig. 3. *SKC with teachers and fellow doctoral students, University of Calcutta, 1959.*

Mahalanobis, left India to settle in the U.S.A. Early in the 1950s, ISI also moved to a new campus far away from the department. This posed difficulty for the department in getting people from ISI as guest teachers.

**Banerjee:** Please tell us about the teachers through whose efforts the department could retain its eminence even after the departure of stalwarts like R. C. Bose and S. N. Roy.

**Chatterjee:** The department had to keep going in its course with only three whole-time teachers P. K. Bose, H. K. Nandi and M. N. Ghosh. P. K. Bose, as the head, had to manage administrative matters, chalk out expansion programs and maintain academic contacts. M. N. Ghosh left the department a few years later. B. N. Ghosh, who had been teaching at Presidency College, joined the department as a whole-timer around 1956, but unfortunately within two years his academic career was cut short by an incapacitating paralytic stroke. Although the department drew upon the expertise of some guest teachers from various institutions and P. K. Banerjee moved in from Presidency College around 1958, the brunt of the responsibility of maintaining the teaching and research activities for the better part of the 1950s and 1960s had to be borne by H. K. Nandi and in this he proved his mettle. Apart from bearing a heavy teaching load and supervising simultaneously the work of several research advisees working in widely different fields, he was the life and soul of Calcutta Statistical Association and the editor of its *Bulletin*. The association had been founded and the *Bulletin* started in the late 1940s to promote the cause of statistics and to provide an outlet for research work carried out particularly in this part of the world.

**Mukerjee:** How did you see the department after you joined there as a faculty?

**Chatterjee:** I joined the department as a lecturer in 1960 after completing my Ph.D. work. Pranab and Jayanta did the same successively in the following two years. After that we lent our hands as far as we could to lighten the workload of Professor Nandi with regard to both the department and the association. Pranab and Jayanta, however, left for the U.S.A. within two to three years. Jayanta returned to join ISI, but Pranab stayed on in the U.S.A. I continued in the department, except for a two-year stint at Lucknow and a one-year visit to Chapel Hill. In the meantime, S. P. Mukherjee and, a little later, B. Adhikari joined the faculty of the department in the mid-1960s and strengthened it in different areas. Another person who enriched the faculty was A. K. Basu; however, he joined much later—early in the 1980s. I do not mention the names of others whose tenure with the department was short, although many of them continued to help the department and the association even while working elsewhere. Incidentally, the department moved to its present spacious location in the southern part of the city in 1964.

**Mukerjee:** How did the department develop in more recent years?

**Chatterjee:** The old guards like P. K. Bose, H. K. Nandi and P. K. Banerjee retired by 1980 and naturally the responsibility of carrying forward their work devolved on some of us. Since the 1980s a new generation of teachers and researchers, some of whom we had helped to develop, has come up and become part and parcel of the department and the association. The baby that was born in 1941 is now a full-grown person standing tall on his feet.

**Banerjee:** How do you evaluate the success of the department in terms of the performance of its alumni in the profession?

**Chatterjee:** The department has every reason to be proud of its distinguished alumni, many of whom have joined the faculties of other prestigious institutions in India and abroad or held responsible positions in various organizations, including the United Nations. Many of them have contributed significantly toward the advancement of the subject. The association, through its bulletin and the various conferences and seminars that it has been helping to organize, has served as a facilitator. One reason why Indian statisticians have distinguished themselves in various fields is that statistics was introduced in India quite early even when the discipline was in its formative stage and a number of first-rate minds were drawn to the subject at that time. The major part of the credit for this must go to Professor Mahalanobis.

## MULTIVARIATE NONPARAMETRICS

**Mukerjee:** On completion of your Ph.D., you started working on multivariate nonparametrics. How did it happen?

**Chatterjee:** In 1962–1963, Professor S. N. Roy visited India, and as was his wont, gave a series of lectures in the department on his current research interests. On this occasion, he spoke on multivariate nonparametric methods. The difficulties of generalizing the univariate nonparametric procedures to the multivariate case are well known. Roy at that time had recently considered, jointly with his Ph.D. student Y. S. Sathe, the problem and, finding that straightforward extension would not work, had adapted the step-down procedure (earlier considered by J. Roy in the parametric context) toward this end. At the end of the series of lectures, Professor Roy conceded that the solution was not fully satisfactory and remarked, "You are welcome to try your hands at it."

**Mukerjee:** You had a significant collaboration with P. K. Sen in this area. Please tell us about this experience.

**Chatterjee:** Until 1962, I was almost innocent of nonparametric theory. Pranab (P. K. Sen), who had been working in that area, was very much conversant with it and said that he had also considered the problem of multivariate generalization earlier and that it looked difficult. We started thinking on the problem together and discussed it off and on between ourselves. It was thus that the principle of conditionally fixing the unordered collection of vectors of variate-wise ranks emerged. Given his familiarity with nonparametric technology, thereafter, Pranab took a leading role and very quickly multivariate versions of the Wilcoxon and median tests for the two-sample location problem were worked out [20]. This was followed by our joint work on tests for equality of association parameters [21] and a multisample version [22] of the first problem. In the meantime, I managed to derive a bivariate extension of the sign test [4].

**Banerjee:** Did the above work have any connection with permutation tests?

**Chatterjee:** The principle of conditionally fixing the unordered collection of rank vectors had some resemblance with that of conditionally fixing the collection of value vectors considered by Wald and Wolfowitz in the context of permutation tests. But permutation tests are not generally accorded the status of nonparametric procedures. In any case, we were unaware of the connection at that time. Pranab and I were very eager to communicate our findings to Professor S. N. Roy, but unfortunately he passed away prematurely before this could be done.

**Banerjee:** Professor Sen left the University of Calcutta in the mid-1960s. What happened next?

**Chatterjee:** Yes, this turned out to be a case of permanent migration to U.S.A. Pranab followed up the development of multivariate nonparametric tests jointly with M. L. Puri and their first book [28] on the topic came out shortly. I too left the department to work at Lucknow University for two years before rejoining in December 1968.

## THE 1970S AND THEREAFTER

**Mukerjee:** You visited the University of North Carolina at Chapel Hill for a year in 1972. Tell us about your work there.

**Chatterjee:** I worked on estimation of the mixture rate of two multivariate populations and an associated classification problem [5, 6]. I also started working with Pranab on semisequential tests for progressively censored experiments [23] and on testing the hypothesis of symmetry for independent but not identically distributed random variables [24]. Incidentally, during that one year Pranab helped me to pick up some of the newer ideas like contiguity and martingales, which were being increasingly used for the development of nonparametrics.

**Banerjee:** The topics you worked on after your return from Chapel Hill were quite diverse. Will you please give us a flavor of this research?

**Chatterjee:** Most of my research during this period 'til the early 1990s was advisee-driven. The topics were diverse because the students whom I advised worked in widely different fields. The development of multivariate nonparametric tests against restricted alternatives was initially one of my main interests in this period. The tests were derived using the union–intersection principle on the basis of the Bahadur slope. A significant challenge involved establishing the power superiority of the tests so developed for restricted alternatives over their unrestricted counterparts when interest lies only in the restricted alternatives. This could be overcome only in the bivariate case [8, 16]. Later, Chinchilli and Sen [25] proved it in some special cases in the multivariate setup. Another problem that I considered in the late 1970s concerned the development of multivariate tolerance sets via density estimation [18].

**Mukerjee:** In the 1980s, you started working also in design of experiments.

**Chatterjee:** At this time, I started teaching advanced experimental design. In the process, I got involved in the investigation of the best response surface design for estimating the optimum point [17] and also in the issue of orthogonality in the case of general asymmetric factorials [7]. Other problems that attracted my attention at this time were those of estimation of variance components in the unbalanced case [15], utilization of multiple scores in nonparametric testing [12], semisequential tests and the change-point problem [11], and the parametric problem of variable selection for multivariate discrimination [19].

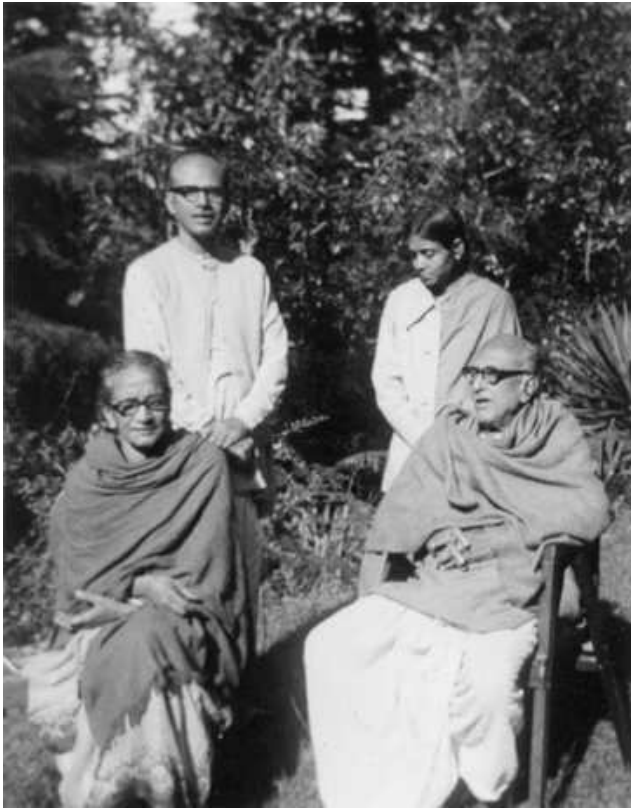

Fig. 4. *SKC with his parents and his wife at a Himalayan hermitage, 1980.*

**Banerjee:** It seems that you were not overtly influenced by the existing research trends. How did you select the specific research problems?

**Chatterjee:** There are teachers who keep themselves posted with the latest developments in the frontier areas. They are able to set topical problems which can be tackled with comparatively less effort. Unfortunately, whether due to my egocentrism or myopic vision, I do not fall into this category. My own tendency has been to select outstanding problems which lay interior to the front line and rely on imagination to formulate and tackle these in different ways.

**Mukerjee:** You became the editor of the *Calcutta Statistical Association Bulletin* in 1979. Tell us about your experience in this capacity.

**Chatterjee:** I took the baton from Professor H. K. Nandi in the relay race that has been going on to keep the *Bulletin* alive and kicking. Nandi steered it for about thirty years, whereas I contributed my mite for fourteen years. I took charge after the *Bulletin* had got well established. While this spared me the pioneer's strain, the responsibility of selecting the publication-worthy papers from among the large number of incoming papers was perhaps a little heavier during my tenure. Also, possibly because of my overperfectionism, I involved myself in the selection process a little more than what editors normally do. This, however, helped to mitigate my own narrowness and egocentrism and made me more familiar with the global developments in statistics.

## FOUNDATIONAL ISSUES

**Banerjee:** Since the late 1980s, you got passionately involved in the foundational issues in statistics. This culminated in the publication of your recent seminal book. How did all these begin?

**Chatterjee:** These issues were troubling me since the mid-1980s. I went through the work of D. Basu and Jack Kiefer. Further discussion with Basu and Jayanta sharpened my discomfiture about the limitation of frequentist procedures [9]. I was not very hopeful about Kiefer's [27] conditionalization, nor was I too comfortable with the Bayesian approach. Jointly with a doctoral advisee, I thought of interpreting the confidence coefficient in a different way by allowing it to depend on the realized data. This permitted detailed statistical inference, which could be developed for two-decision and multiple-decision problems [13, 14].

**Banerjee:** Specifically, what made you undertake the project of writing the book *Statistical Thought: A Perspective and History* [10] on the foundations of statistics?

**Chatterjee:** Until even the early 1990s, I was groping to see whether non-Bayesian inference procedures such as those based on the conditional approach or the use of *p*-values could be established on a firmer basis. The work on detailed statistical inference was an offshoot of this endeavor. But none of these appeared to lead to a comprehensive resolution of the difficulty. From 1992 onward, I was feeling that my association with statistics for almost forty years could end in a thankless consummation if I quit leaving all the threads with dangling loose ends. This was the principal motivation for writing a book on the foundations of statistical thought and their evolution. Also, as I had been teaching the history of statistical thought for almost fifteen years, closer examination of the conflicts among different schools of statistical inference naturally seemed to be in order.

**Mukerjee:** How did this project help you in resolving the dilemma?

**Chatterjee:** Upon undertaking the project, I realized that any satisfactory resolution of the dilemma would require embedding the whole corpus of statistical inference within a wider philosophical canvas. As I read various books on philosophical induction, the idea crystallized in me that each of the different approaches to statistical inference is perfectly natural in its appropriate setting. Also, I could see that the different approaches arose because different conceptions of probability were invoked at different stages of the inferential process. Writing the book itself was an education to me—many blank patches in my thought world got filled up in the process. I do not expect that my resolution of the dilemma will satisfy every member of every school. However, subjectively, I myself am satisfied and can call it a day with a clear conscience.

**Banerjee:** In the book, you discuss a counterexample to the widespread belief that the principles of sufficiency and conditionality imply the likelihood principle. Please elaborate on it and its implications.

**Chatterjee:** The weakness of the proof of the implication that you are talking about had been noted earlier in abstract terms, for instance, by Durbin [26]. I tried to make it more explicit in a general setting in my Science Congress presidential address [9]. While writing the book, I thought that the point would become even clearer if stated in terms of direct and inverse binomial sampling. In the first draft I had put it in the form of a footnote. Sir David Cox, who saw the draft, prevailed upon me to incorporate it in the body of the text.

**Mukerjee:** How long did the project take? Also, did you receive any support from anyone in this endeavor?

**Chatterjee:** I started writing the book in 1997 and finished it in five and-a-half years. Sir David Cox magnanimously agreed to review the draft of the first four chapters and was kind enough to send his encouraging comments.

**Mukerjee:** You conclude the book with a strong plea to statisticians to shun dogmatism and be eclectic. Is it because of your realization as you had just talked about or is it because of your concern that disagreement among statisticians may create an adverse professional image in the scientific community, including the users?

**Chatterjee:** Actually, I read one paper where it was stated that this kind of squabbling among the statisticians is affecting their credibility to outsiders. Fortunately, however, practical conclusions usually turn out to be the same, whatever theoretical approach one follows.

**Banerjee:** A reviewer of your book writes about a strong Anglo-Indian bias in presentation, in particular, regarding developments during the twentieth century. Is it inadvertent?

**Chatterjee:** Frankly, I don't understand what is meant by an Anglo-Indian bias. However, I deliberately gave some references to Indian authors which are possibly not well known but which, I am convinced, deserve to be cited for the sake of academic justice.

## BE THOU AT REST

**Mukerjee:** Throughout, you have been extremely parsimonious in publishing your research. We know that you cared to publish only those results that passed your own very stringent screening. This is much in contrast with the idea of "publish or perish" which is now quite common in some quarters. Will you please elaborate?

**Chatterjee:** I cannot give any satisfactory answer to this except saying that I have perhaps been naturally so made up. I do neither have the dynamism nor the passion for research possessed by some of my friends. Also, perhaps I was somewhat influenced by Professor H. K. Nandi's philosophy of giving more importance to "being" than "achieving." Incidentally, I may mention that a famous biochemist (I forget his name, but he was a co-discoverer of streptomycin) visited Calcutta some thirty years back. In the course of addressing a group of students, he gave a piece of advice, which struck a chord in my mind: "Always have two frames of reference in your life—one short-term and one long-term frame." Most people, except possibly geniuses, start their lives with one short-term frame, follow it up with another and the sequence goes on. Very often, these short-term frames quickly exhaust their relevance. I believe, as one grows in years, one should choose and stick to a long-term frame of reference in life. I think, around the mid-1970s, I was fortunate to find such a long-term frame of reference. Whatever little I did or did not thereafter can be explained in the light of that.

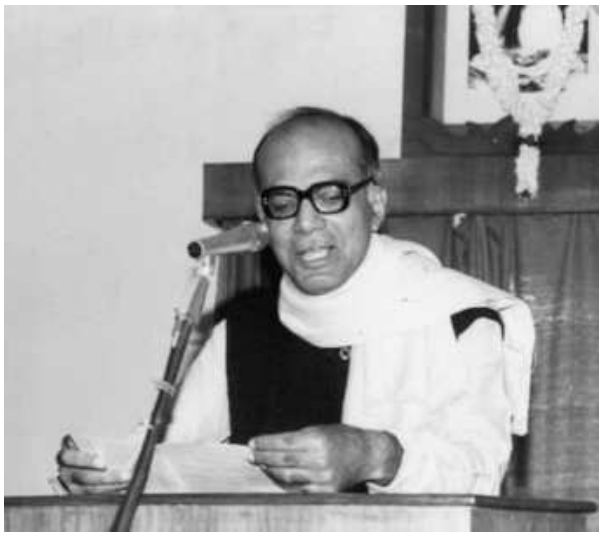

Fig. 5.  *SKC at a departmental seminar, 1984.*

**Banerjee:** Besides research, you are considered to be one of the finest teachers of our discipline. Your lucid exposition of such diverse areas as nonparametric statistics, decision theory and design of experiments stimulated many of us to pursue doctoral and subsequent work in these areas. In view of the premium given in today's academic world on research alone, do you have any reason to think that spending so much time on teaching is worthwhile? For example, did it hinder your own research in any way?

**Chatterjee:** As you have noticed, my own research during the latter part of my academic career, except my investigation into the foundational issues, was mostly advisee-driven. There was no conflict between this and teaching. There was also no conflict between teaching and my work on foundations, since one of the subjects then being taught by me was the history of statistical thought. Besides, as I said earlier, one should assess the importance of different duties in one's life in terms of one's long-term frame of reference. Judged in these terms, there can be no conflict between teaching and research.

**Mukerjee:** While many of your contemporaries moved out for greener pastures, you decided to stay back here even though you were offered lucrative positions elsewhere (we heard about that). What made you stay here?

**Chatterjee:** Firstly, I was perhaps not very ambitious in the usual sense. Secondly, I had a very favorable ambience in my place of work—appreciative superiors and cooperative and friendly colleagues and juniors, including students. Thirdly, the members of my family, particularly my wife, were not at all demanding and were satisfied with whatever they had. Possibly all these led me to stay where I was during the last thirty years or so of my working life.

**Mukerjee:** In continuation of the last question, did the traditional Indian system of human values affect your decision in any significant way?

**Chatterjee:** Not really, because people who remain content with their present state and go on performing their work wherever they are, are basically alike all over the world in all ages. In this context, I recall one saying of Caliph Ali [Caliph Ali (600–661) was the fourth caliph of Islam] which Sister Nivedita [Sister Nivedita (1867–1911), a well-known writer, was Irish by birth. Upon becoming a disciple of Swami Vivekananda, she came to India and spent the rest of her life working for social and spiritual causes.] quoted at the beginning of one of her books: "Be thou at rest from seeking thy place in life, for thy place in life is seeking after thee". As I grow in years, I more and more realize the truth of this statement.

## TOWARD THE FUTURE

**Banerjee:** This concerns theory versus practice. Do you see a conflict between the two?

**Chatterjee:** Research problems in statistics in their original form—I am excluding derived problems—always have reference to an empirical entity that looms in the background. This is the main difference between a research problem in statistics and one in pure mathematics. The principal objective of training a statistician should be to inculcate in the person a data sense; that is, a statistician should be instinctively able to replace the empirical entity at the back by numbers representing its various features. For developing this data sense, both theoretical studies and exposure to practice are necessary. Once somebody develops this data sense, he or she can pursue theoretical research or work in a statistical office, depending on his or her situation.

**Mukerjee:** Please say a few words on statistics teaching and research in today's changing world.

**Chatterjee:** The computer revolution will have a profound impact. For example, models represented by software may progressively replace mathematical models. Thus the size of a test or the confidence level of a confidence interval may be derived increasingly through simulation, instead of being mathematically deduced from a model. Ultimately, a conclusion, such as one to reject a model, may be equally convincing whether the degree of assurance is deduced mathematically from the model or via simulation.

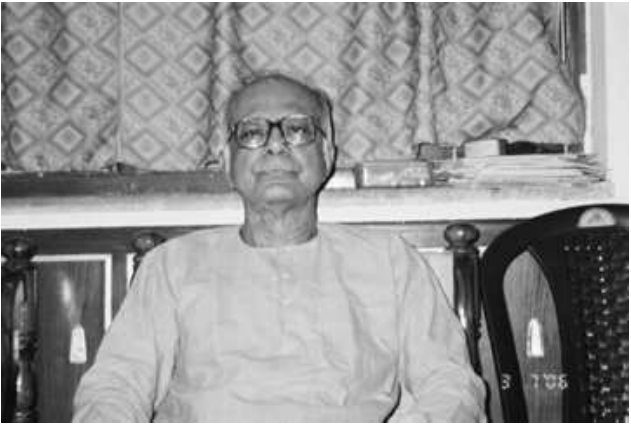

Fig. 6. *Shoutir Kishore Chatterjee, July 2006.*

**Mukerjee:** Do you foresee any limitation of statistics as an instrument of knowledge?

**Chatterjee:** We noted earlier that statistics works by replacing an entity of the empirical world by numerical observations. As instrumentation develops further and further, and one may reach a stage when simultaneous observation of more than one feature may be impossible. This has happened in quantum physics. I have apprehension that this may happen in genetics too. We hear so much about gene mapping, but as far as I know, the activation of particular genes in an organism is indeterminate. If this happens, what role can statistics have in that stage?

**Banerjee:** Over the last three years or so, you maintained a significant interest in the human development index. How were you motivated to work on this topic?

**Chatterjee:** For a long time it was taken for granted that the economic development of a country and the development of its people are synonymous. In recent years, emphasis has been put on human aspects of development as reflected in the human development ment reports of the United Nations Development Program. But this is being done in an ad hoc manner without formulating any comprehensive frame of reference. For many years, I have been a student of Swami Vivekananda's [Swami Vivekananda (1863–1902) was a major exponent of Vedanta philosophy and the founder of the world-wide Ramakrishna movement. In 1893, he went to the U.S.A. to represent Hinduism in the World Parliament of Religions in Chicago.] writings and I have felt that these contain the seed of such a comprehensive framework from the Vedantic (Vedanta is one of the six classical systems of Indian philosophy) standpoint. I have been working to make this explicit for the last few years. But I do not know whether it will be given to me to develop this fully. In any case, this pursuit keeps me engaged in the study of the literature on human development and induces me to make a thorough study of Vivekananda's work and other relevant material.

**Mukerjee:** What are your hobbies and other interests?

**Chatterjee:** Mainly reading, particularly philosophical literature. I theoretically believe that at a certain stage of life one should try to interiorize one's existence, that is, one should not depend too much on outside things for one's mental sustenance. Of course, it is very difficult to put this into practice, particularly for people who have been preoccupied with outside work or interaction with other people all their life.

**Banerjee:** And finally, what is your advice to the younger generation of researchers?

**Chatterjee:** For theoretical researchers, integrity and sincerity are the two primary requisites. Personally, I have seen that imagination or intuition, bridled by reason, helps in the solution of difficult problems. Furthermore, although initially it is difficult to practice, some degree of detachment is essential for success in research in the long run.

**Banerjee and Mukerjee:** As always, it was most inspiring to talk to you. Thank you very much for the privilege of this interview.

## ACKNOWLEDGMENT

Tathagata Banerjee and Rahul Mukerjee thank Dr. Gourangadeb Chattopadhyay, University of Calcutta, for his help in procuring some of the photographs.